\documentclass[11pt, oneside]{article}   	
\usepackage[margin=1in]{geometry}               
\usepackage{graphicx}				
										
\usepackage{amssymb}
\usepackage{physics}
\usepackage{amsmath}
\usepackage[dvips]{epsfig}

\usepackage{amsfonts}
\usepackage{subfig}

\usepackage{mathrsfs}

\usepackage{bm}

\usepackage[title]{appendix}

\usepackage{cite}

\usepackage{authblk}

\def\cchi{\raise2pt\hbox{$\chi$}} 

\newcommand\bsdot{\ensuremath{\boldsymbol{.}}}

\title{\bf{Quantum Lattice Representation for the Curl Equations of Maxwell Equations}}

\author{George Vahala}
\author{John Hawthorne}
\affil{Department of Physics, William \& Mary, Williamsburg, VA23185}
\author{Linda Vahala}
\affil{Department of Electrical \& Computer Engineering, Old Dominion University, Norfolk, VA 12319}
\author{Abhay K. Ram}
\affil{Plasma Science and Fusion Center, MIT, Cambridge, MA 02139} 
\author{Min Soe}
\affil{Department of Mathematics  and Physical Sciences, Rogers State University, Claremore, OK 74017}

\begin{document}
\maketitle
$\bf{Abstract}$:  A quantum lattice representation (QLA) is devised for the initial value problem of one-dimensional (1D) propagation of an electromagnetic disturbance in a scalar dielectric medium satisfying directly only the two curl equations of Maxwell.  It si found that only 4 qubits/node are required.  The collision, streaming, and potential operators are determined so as to recover the two curl equations to second order.  Both polarizations are considered.

\section{Introduction}
\quad In plasma physics it is not uncommon to truncate the full Maxwell equations by dropping enforcement of the two divergence equations $\nabla \bsdot (\epsilon \mathbf{E})  \quad=  \rho,  \quad  \nabla \bsdot \mathbf{B} \ = \ 0$, treating them as initial conditions.  The remaining two curl equations are
\begin{align}
\nabla \times \mathbf{E} \ &= \ - \frac{\partial \mathbf{B}}{\partial t}  \ & 
\ \nabla \times \mathbf{B} \ &= \mu_0 \mathbf{J} + \mu_0 \frac{\partial \, \epsilon \, \mathbf{E}}{\partial t} .
\label{fam}
\end{align}
Here $\mathbf{E}$ and $\mathbf{B}$ are the electric and magnetic fields, $\rho$ is the free charge density, $\mathbf{J}$ is the free current density,  $\mu_0$ is the (constant) magnetic permeability, and $\epsilon$ is the (scalar) dielectric function.  Of course, if $\nabla \bsdot \mathbf{B} \ = \ 0$ initially, then theoretically it remains zero for all time.  Unfortunately, numerical codes will wobble in their time evolution of $\nabla \bsdot \mathbf{B}$ and thus require divergence cleaning or the introduction of a vector potential.  Gauss' law can also be problematic and not so easily discarded.  Here we shall consider one dimensional (1D) propagation of an electromagnetic pulse, governed by the two curl Maxwell equations of Eq, (1).  In a scalar dielectric medium these 1D pulse are transverse so that automatically $\nabla \bsdot (\epsilon \mathbf{E}) = \nabla \bsdot \mathbf{B} \ = \ 0$ for all times.

Recently [1-5] we have been developing quantum lattice algorithms (QLA) for the solution of the full Maxwell equations in 1D, 2D and 3D.  In this note we wish to develop a 1D QLA for the solution of the two curl Maxwell equations.  This has been motivated by a conference paper simulation by Jestadt, Appel and Rubio [6] in which they also restrict their attention to the two curl Maxwell equations.  Following the work of Khan [7] and the introduction of the Riemann-Silberstein-Weber vector 
\begin{equation}
\label{R-S vector}
\mathbf{F^{\pm}} = \sqrt{\epsilon} \mathbf{E}  \pm i \frac{\mathbf{B}}{\sqrt{\mu_0}}.
\end{equation}
 they [6] reduced their system to a 6-spinor Schrodinger-like equation.  The exponential evolution operator is then approximated using the standard Baker-Campbell-Hausdorff expansion.  This introduces both a first and second order derivative of the refractive index $n(\mathbf{x}) = \sqrt{\epsilon(\mathbf{x}})$.  Their resulting simulation of a 1D wave packet incident from a vacuum into a dielectric medium yields a transmitted wave packet with similar wavelength to the incident wave packet.  However, one expects the transmitted wavelength to be reduced by the corresponding ratio of the media refractive indices.
 
 In Sec. 2 we shall develop a 1D QLA for the two curl equations of Maxwell based on the Khan representation that draws a strong analogy with the Dirac equation for a massless particle [8-10].  The QLA [1-5, 11-12] consists of a sequence of non-commuting collide-stream operators and some potential collision operators so chosen that in the continuum limit one recovers the Khan representation.  In Sec 3 we present initial value 1D QLA simulations for a Gaussian pulse propagating in the x-direction and consider the two polarizations:  one with non-zero field components $E_y$ and $B_z$ and the other with non-zero $-E_z$ and $B_y$.  The pulse propagates from a region of refractive index $n_1$ that is joined to another dielectric of refractive index $n_2$ by a thin boundary layer.
   We recover exactly the same physics as with the QLA for the full Maxwell equations [1,2]:  in particular the transmitted to initial field amplitude scales as $2 \sqrt{n_1 n_2}/(n_1+n_2)$  - which is an extension of the standard boundary value result [13] for a plane electromagnetic wave by the factor $\sqrt{n_2/n_1}$.  We also consider the reflection and scattering of a Gaussian wave packet and indeed see a significant change in the wavelength of the transmitted packet.  Only 4-qubits per spatial grid point are required for this new QLA.  Finally, in Sec. 4 we present some conclusions and comments on further work.
 
 \section{1D QLA for the two curl equations of Maxwell}
 
 \quad  It is convenient to introduce the qubits as just the Riemann-Silberstein-Weber vectors, Eq. (2).  With that choice we now proceed to determine the required form of the collision, scattering and potential operators that will recover the two curl equations of Maxwell in the continuum limit.  As details are presented in our earlier papers [1-5], we just summarize the results here.    Since we are considering 1D propagation in the x-direction, we automatically have $E_x =0= B_x$.  This reduces the number of qubits/lattice site to 4:
 \begin{align}
 q_1 = n(x) E_y + i B_y  , \qquad    q_2 = n(x) E_z + i B_z  \\
 q_4 = n(x) E_y - i B_y  , \qquad     q_5 = n(x) E_z - i B_z ,
 \end{align}
 since $q_0 = n(x) E_x + i B_x = 0 = q_3 = n(x) E_x - i B_x $.
An appropriate collision matrix acting on the 4-spinor $Q = (q_1 \quad q_2  \quad q_4  \quad q_5)^T$ is the $4 \cross 4$ unitary matrix
 \begin{equation}
C(\theta) = 
  \begin{bmatrix}
  \cos \theta & i \sin \theta & 0 & 0   \\
  i \sin \theta  &  \cos \theta  &  0  &    0 \\
    0 & 0  & \cos \theta  &  -  i \sin \theta  \\
    0  &  0  & -i \sin \theta  &    \cos \theta  \\
   \end{bmatrix}.
\end{equation}
and the interleaved non-commuting sequence of collide-stream operators
\begin{equation}
U= S^-_{25}.C^+.S^+_{25}.C.S^+_{14}.C^+.S^-_{14}.C.\quad  S^+_{25}.C.S^-_{25}.C^+.S^-_{14}.C.S^+_{14}.C^+
\end{equation}
 where $S_{14}^+$ streams qubits $q_1$ and $q_4$ one lattice unit to the right while not streaming qubits $q_2$ and $q_5$.  $C^+$ is the adjoint of the unitary matrix $C$.  To perturbatively recover the two curl equations of Maxwell one will need to introduce the potential operators
 \begin{equation}
 V_1(\alpha) = 
  \begin{bmatrix}
  \cos \alpha & - \sin \alpha & 0 & 0   \\
    \sin \alpha   &  \cos \alpha  & 0  &    0 \\
    0 & 0  & \cos \alpha  &  \sin \alpha  \\
    0  &  0  & - \sin \alpha  &    \cos \alpha  \\
   \end{bmatrix}.
\end{equation}
and 
 \begin{equation}
 V_2(\alpha) = 
  \begin{bmatrix}
  \cos \alpha & 0 & 0 & - \sin \alpha   \\
     0  &  \cos \alpha  & \sin \alpha   &    0 \\
    0 & \sin \alpha  &  \cos \alpha  &  0 \\
    - \sin \alpha  &  0  & 0  &    \cos \alpha  \\
   \end{bmatrix}.
\end{equation}
and the rotation angles  $\theta = -\epsilon /4 n(x)$ and $\alpha = i \, \epsilon^2 n'(x)/2 n^2(x)$.  The perturbation parameter $\epsilon$ will turn out to be the speed of light in that medium.  The QLA time evolution of the 4-qubits is then determined by
\begin{equation}
  Q(t+\delta t) = V_2(\alpha) V_1(\alpha) .  U  Q(t)
\end{equation}
In the continuum limit the above QLA reduces, to $\O(\epsilon^2)$,
 \begin{equation}
\label{R_S InHomox1}
\frac{\partial}{\partial t}
\begin{bmatrix}
q_1   \\
q_2  \\
q_4   \\
q_5   \\
\end{bmatrix}  	
= - \frac{i}{n(x)} 
\frac{\partial}{\partial x}
\begin{bmatrix}
-q_2  \\
q_1  \\
q_5  \\
-q_4  \\  
\end{bmatrix}
+ \frac{i \, n'(x)}{2n^2(x)}
\begin{bmatrix}
-q_2 - q_5  \\
q_1 + q_4  \\
q_2 + q_5  \\
-q_1 - q_4  \\  
\end{bmatrix}
\end{equation} 
 From  Eqs. (3) and (4), it is readily shown that the continuum Eq. (10) are nothing but the two curl equations of Maxwell  for both polarizations.

 \section{1D QLA simulations of the two curl equations of Maxwell}
 \subsection{Gaussian Pulse}
 \quad  Consider a Gaussian pulse propagating from a vacuum (with refractive index $n_1 = 1$) into a medium with refractive index $n_2 = 2.5$.  There is a sharp boundary layer of approximate thickness of 10 lattice units connecting $n_1$ to $n_2$, Fig. 1.
 \begin{figure}[!h!p!b!t] \ 
\begin{center}
\includegraphics[width=5.1in]{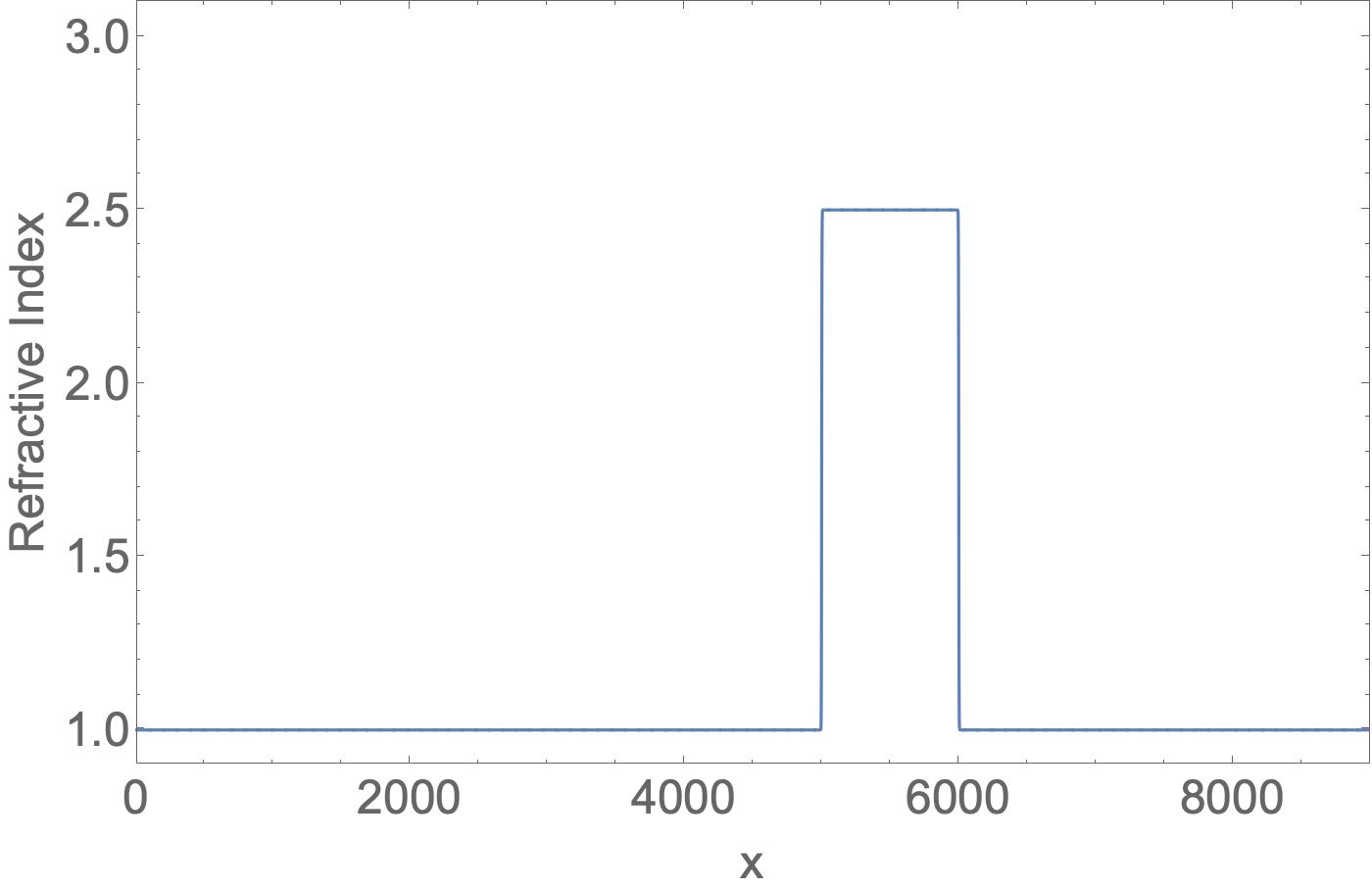}
\caption{The refractive index of the medium:  vacuum for $0<x<5000$ and $6000<x<9000$ with dielectric in region $5000 < x < 6000$  with $n_2	=2.5$.  The boundary layer is about 10 lattice units thick.
}
\end{center}
\end{figure}
 
Figure 2 shows the pulse of width 600 lattice units with normalization such that its speed of propagation is $\epsilon=0.3$, and the magnitude of the electric field component is equal to its magnetic field component.          
  \begin{figure}[!h!p!b!t] \ 
\includegraphics[width=5.1in]{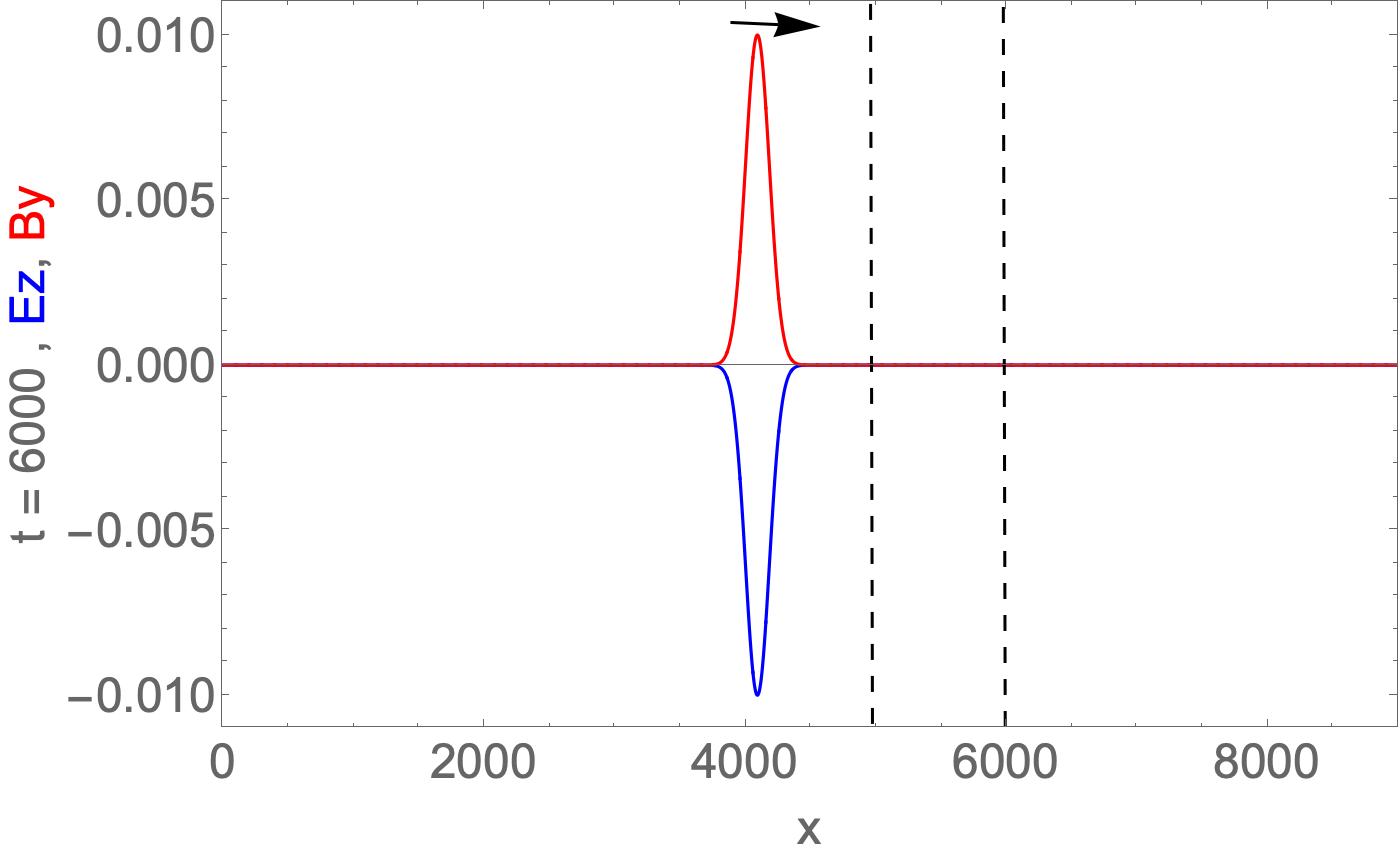}
\begin{center}
 (a)   polarization $E_z$ 
\end{center}
\includegraphics[width=5.1in]{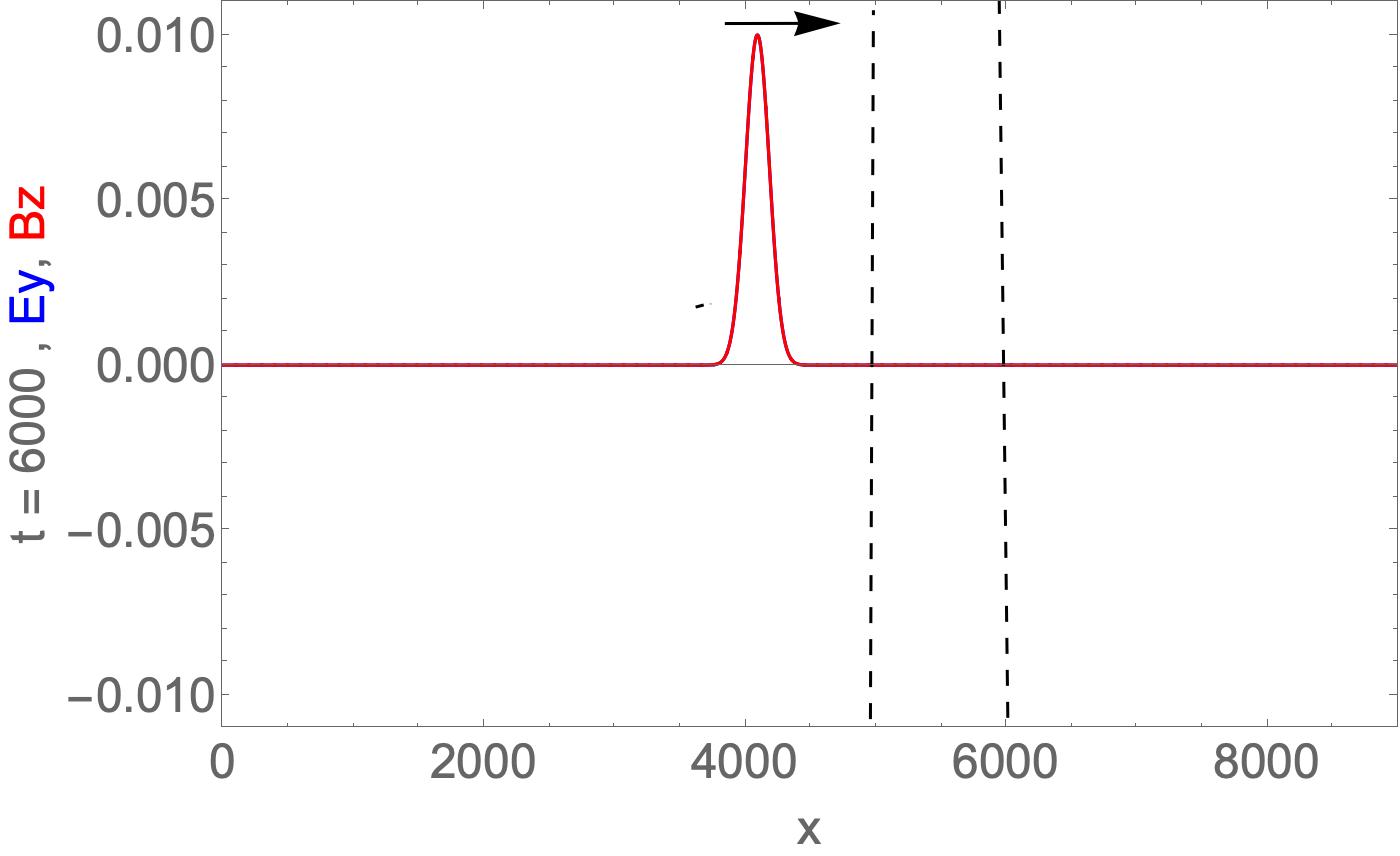}
\begin{center}
(b)  polarization  $E_y$
\end{center}
\caption{The propagation of a Gaussian pulse at time t = 6000 for polarization with non-zero field components (a) $E_z < 0$ (in blue) , $B_y > 0$ (in red) ;  (b)  $E_y >0$ (in blue) , $B_z > 0$ (In red).  In our normalized units, for case (b),  $E_y = B_z$ and the two profiles overlay each other.
}
\end{figure}

By t = 12000, the Gaussian pulse has encountered its first reflection/transmission.  Since the propagation is from low to higher refractive index , the reflected electric field component at the vacuum-dielectric interface undergoes a $\pi$ phase reversal, as seen in Fig. 3.  (This, of course, agrees with the standard plane wave boundary value result [13]).

  \begin{figure}[!h!p!b!t] \ 
\includegraphics[width=5.1in]{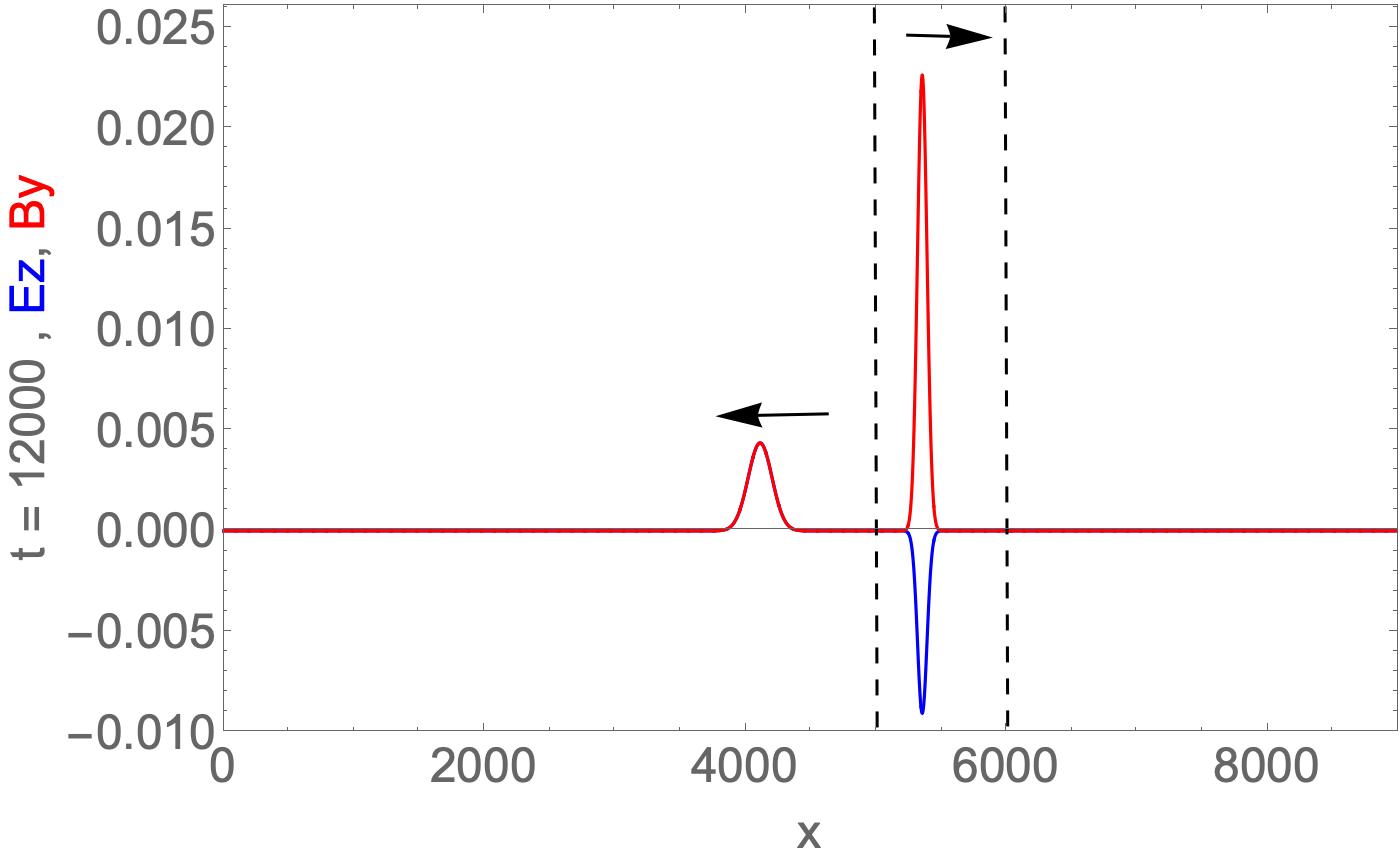}
\begin{center}
(a)   polarization $E_z$
\end{center}
\includegraphics[width=5.1in]{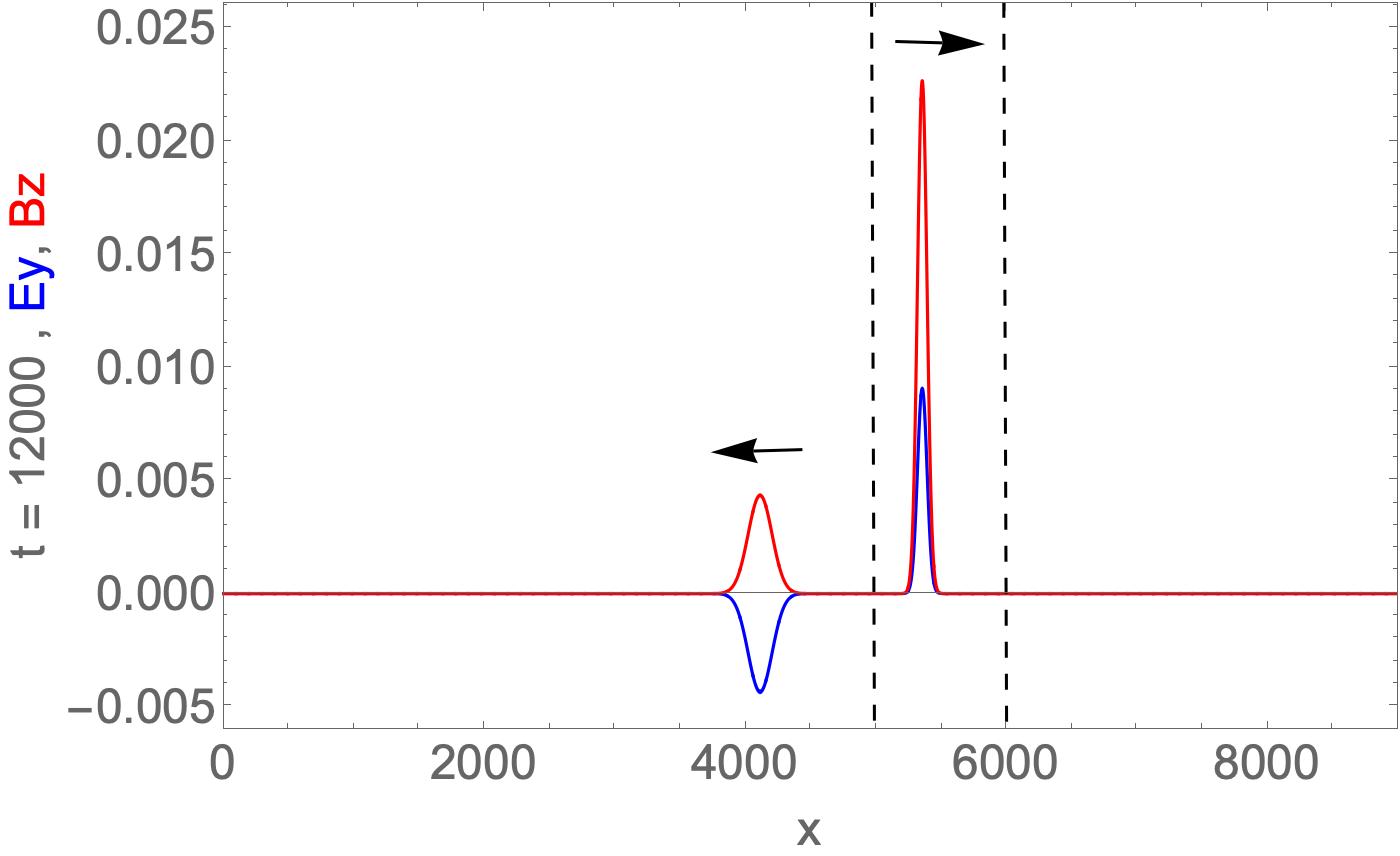}
\begin{center}
 (b)  polarization  $E_y$
 \end{center}
\caption{The propagation of a Gaussian pulse at time t = 12000 after the first encounter with the dielectric slab around $x=5000$.  The pulse width and speed in the dielectric are reduced by the facctor $n_2/n_1$. (a) the reflected $E_z$ ( blue) changes its phase by $\pi$ and becomes positive and overlays $B_y$ (red) while the transmitted pulse has these componentsout of phase;  (b)  the reflected pulse now has the $E_y$ (blue) and $B_z$ (red) out of phase while the transmitted pulse has these components in phase.
}
\end{figure}

Finally we plot the pulses after their interaction with the backside of the $n_2$ dielectric at $x=6000$, Fig. 4.  Since the pulse now is propagating from higher to lower refractive index,  $n_2 > n_1$, it is now the reflected magnetic field that undergoes a $\pi$-phase reversal.

  \begin{figure}[!h!p!b!t] \ 
\includegraphics[width=5.1in]{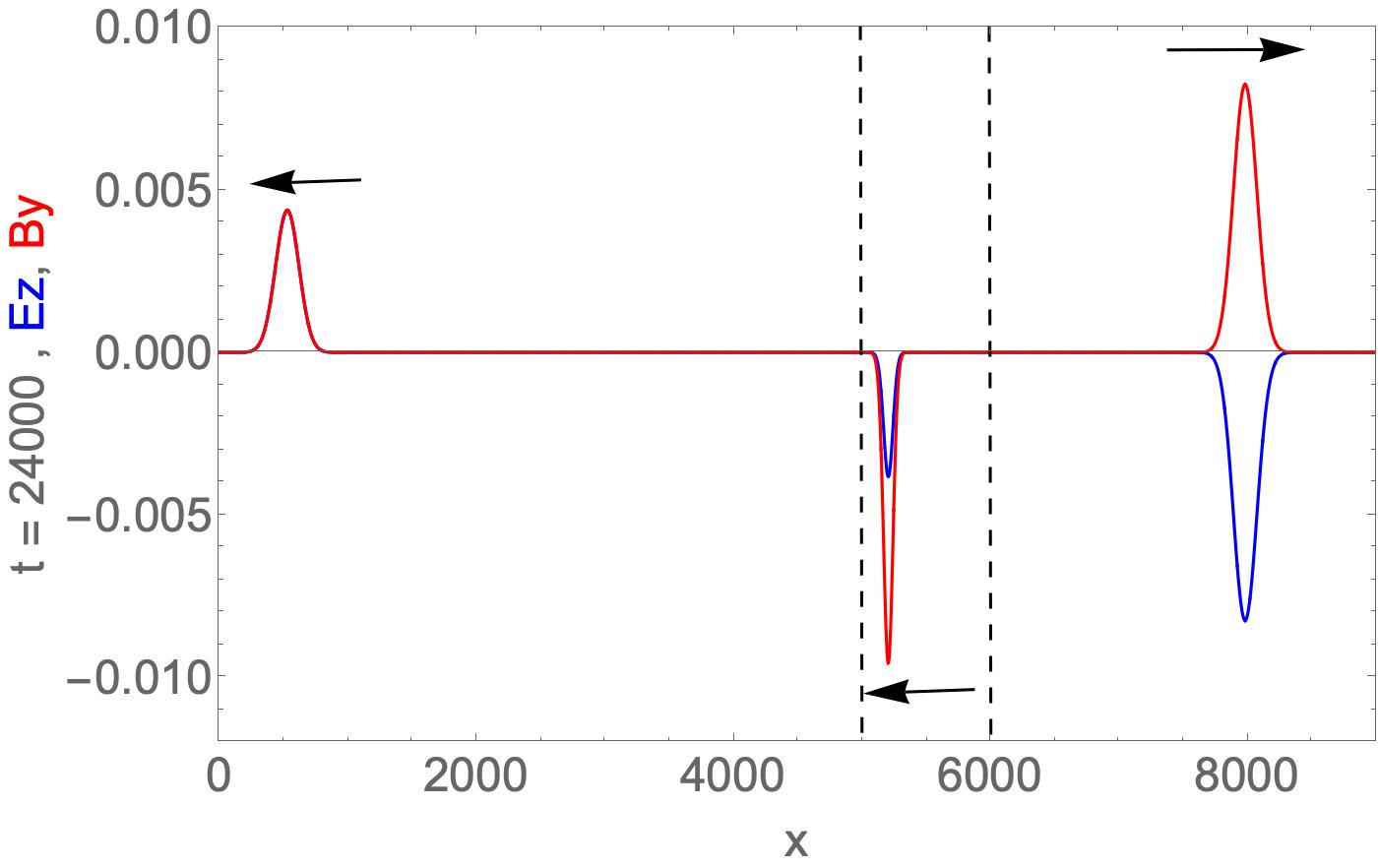}
\begin{center}
 (a)   polarization $E_z$ 
\end{center}
\includegraphics[width=5.1in]{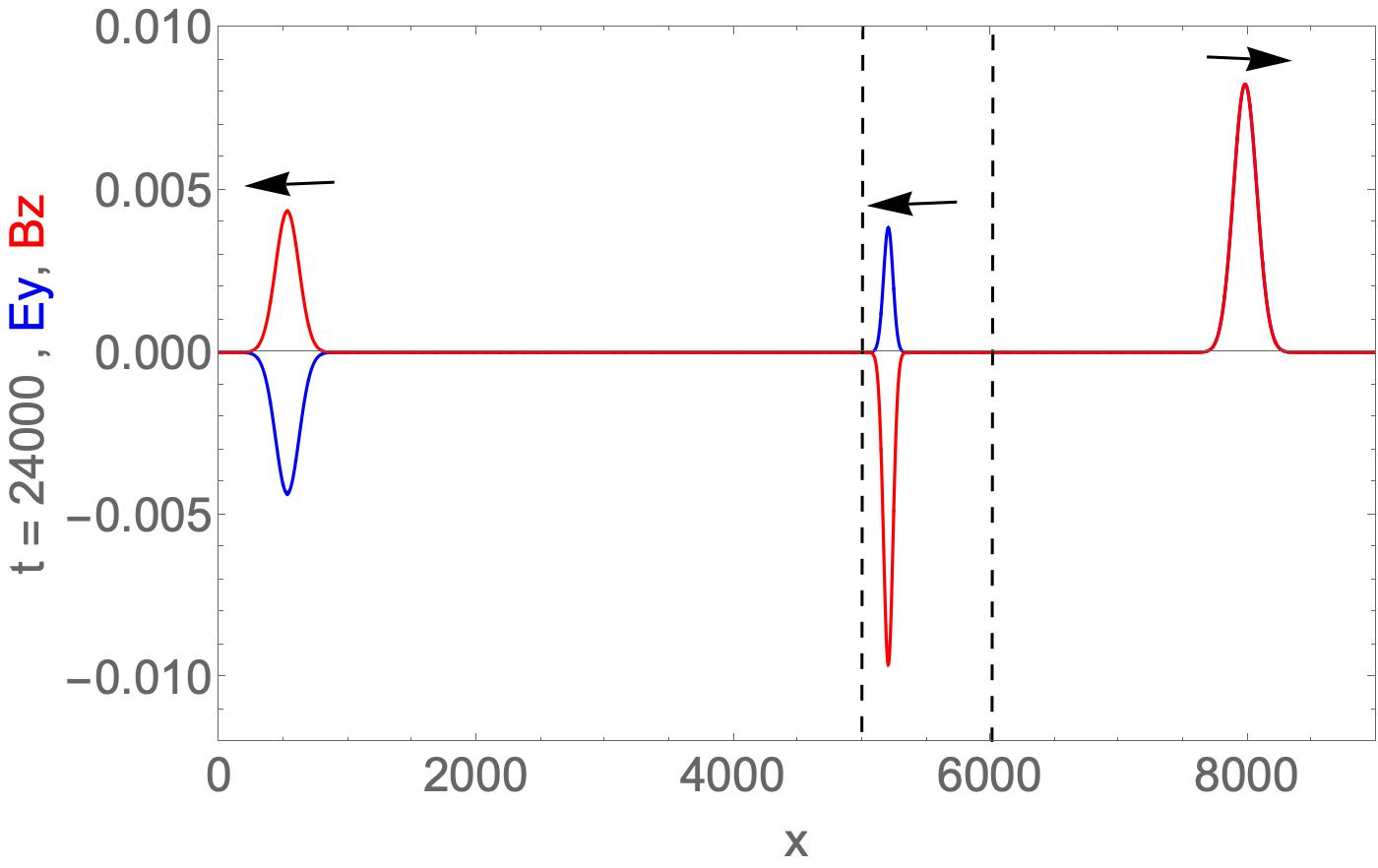}
\begin{center}
(b)  polarization  $E_y$
 \end{center}
\caption{The propagation of a Gaussian pulse at time t = 24000 after the seccond encounter with the dielectric slab-vacuum interface, but now at $x=6000$.  The pulse width and speed in the dielectric are reduced by the factor $n_2/n_1$. (a) the reflected $B_y$ (red) changes its phase by $\pi$;  (b)  the reflected pulse now has the $E_y$ (blue) and $B_z$ (red) out of phase while the transmitted pulse has these components in phase.
}
\end{figure}

In our earlier 1D  QLA for the full set of Maxwell equations, we showed and developed a theory that for normal incidence the electric field amplitude of the transmitted to incident pulse is augmented by a factor of $\sqrt{n_2/n_1} $ over the plane wave boundary value result of $2 n_1/(n_1 + n_2)$. For reflection/transmission at the first interface, we find that both polarizations yield $|E_{trans}/E_{inc} | = 0.91$, Fig. 3.  For the subsequent  reflection/transmission at the back face of the dielectric slab, we find the $E_z$ polarization  that $|E_{trans}/E_{inc} | = 0.88$, Fig. 4a.  For the $E_y$ polarization, the ratio $|E_{trans}/E_{inc} | = 0.89$, Fig. 4b.  The theoretical initial value problems result for $n_1=1, n_2=2.5$ is 0.90 .  (The simple standard boundary value for plane waves, however,  yields a ratio of $0.57$.)

\subsection{Gaussian wave packet}
We plot the electric field $E_y$ for the Gaussian wave packet as it propagates towards the dielectric slab which starts around x = 5000, Fig 5. 
  \begin{figure}[!h!p!b!t] \ 
\begin{center}
\includegraphics[width=5.1in]{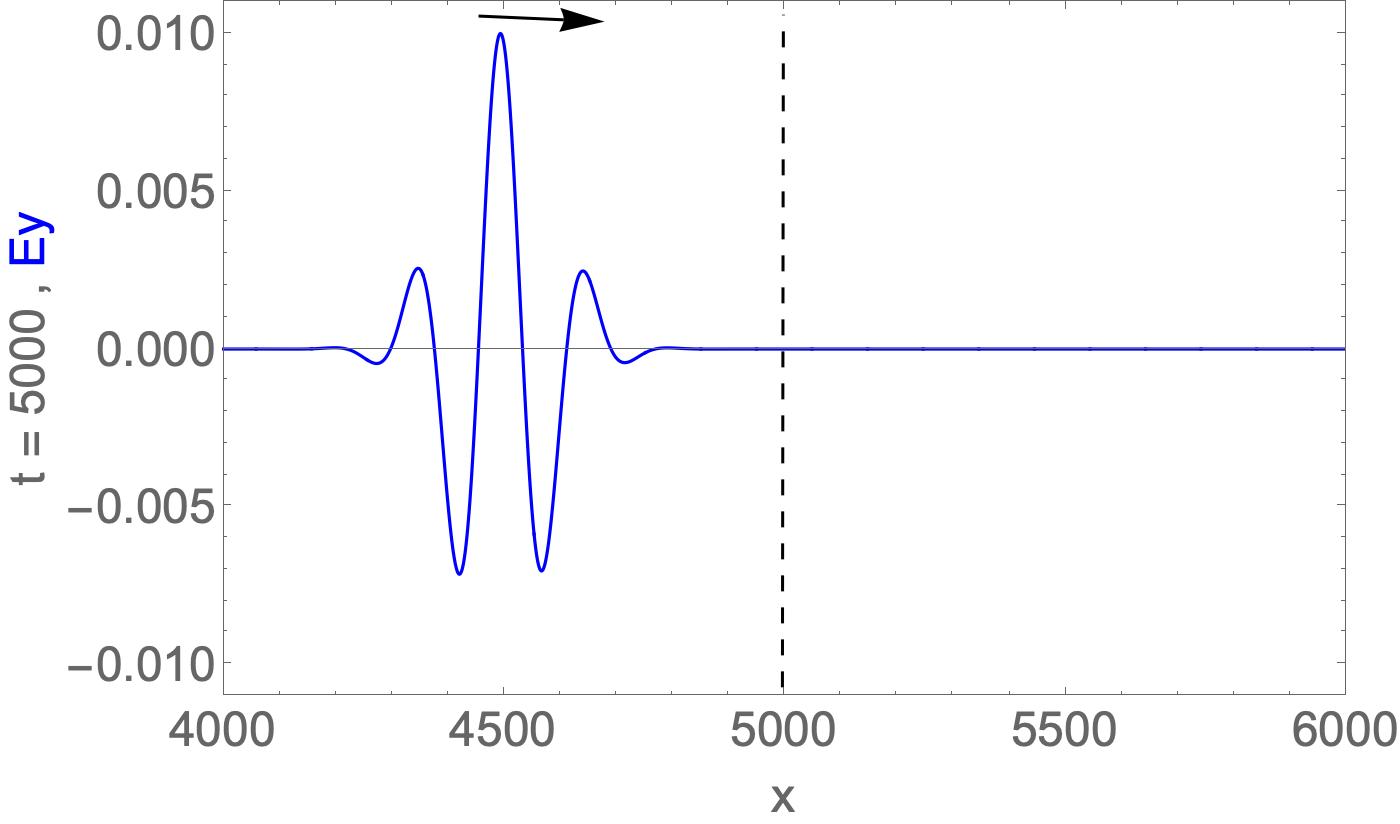} 
\caption{The electric field $E_y$ of a Gaussian wave packet in medium $n_1=1$ as it propagates towards a dielectric slab $5000 < x < 6000$ with $n_2=2$.
}
\end{center}
\end{figure}
After the first reflection/transmission at $x=5000$ one notices the $n_1/n_2$ reduction in the wavelength of the transmitted wave packet within the dielectric $n_2$.  The reflected electric field of the wave packet undergoes a $\pi=$phase change as expected from plane wave boundary value theory.  However the relative field amplitudes do not seem to follow a simple mathematical expression - clearly nothing like that for a boundary value plane wave or an intial value pulse.   
Partly this could arise from the importance of the normalization constraint in the Gaussian pulse theory [2]: with oscillations in the profile this constraint becomes much less effective. Interestingly, there is a slight break in symmetry in the transmitted electric field - as seen in Fig. 6
  \begin{figure}[!h!p!b!t] \ 
\begin{center}
\includegraphics[width=5.1in]{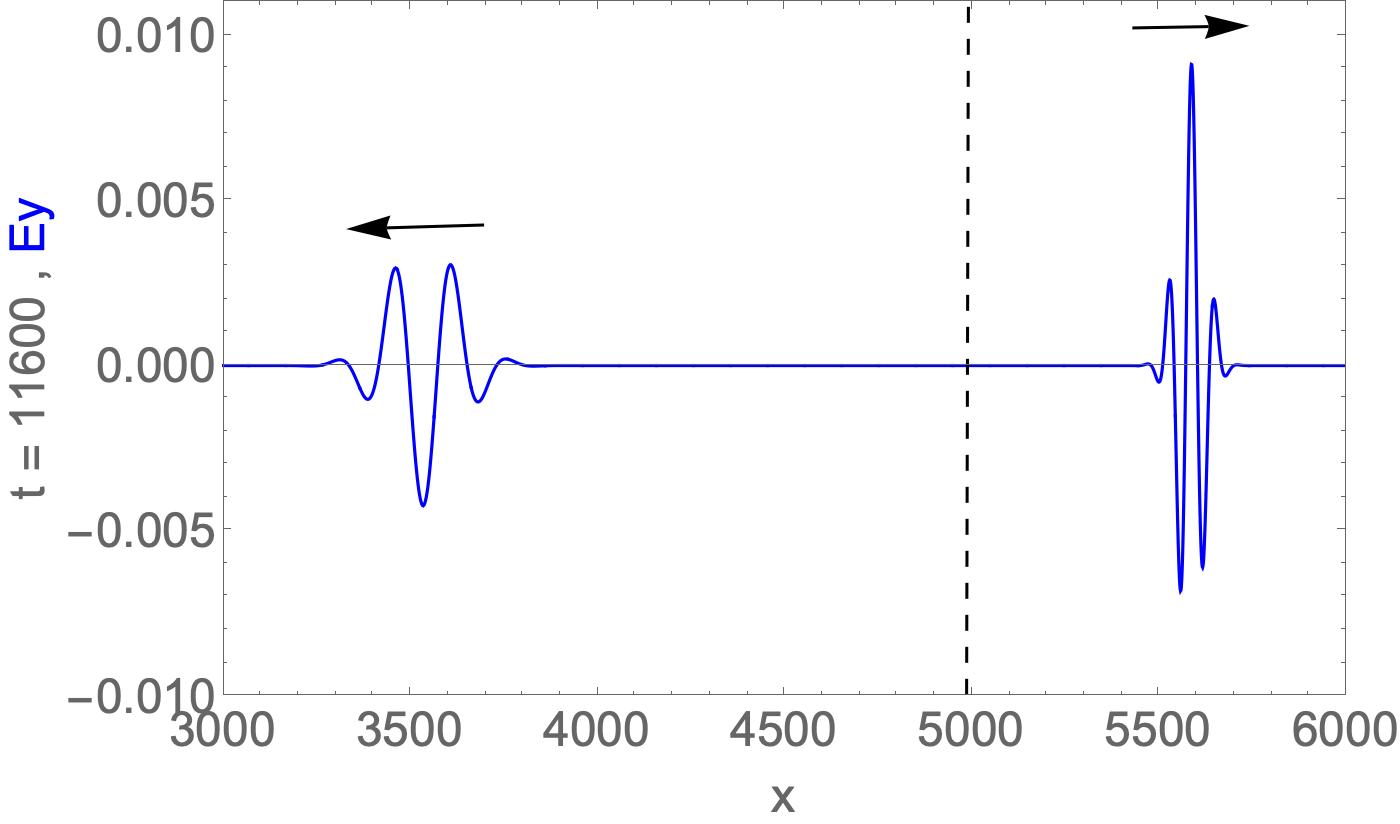} 
\caption{The electric field $E_y$ of a Gaussian wave packet following the first reflection/transmission at the interface $x=5000$.
}
\end{center}
\end{figure}

Finally in Fig 7 we plot the electric field of the wave packet following the second reflection/transmission at the back of the dielectric slab at $x=6000$.
  \begin{figure}[!h!p!b!t] \ 
\begin{center}
\includegraphics[width=5.1in]{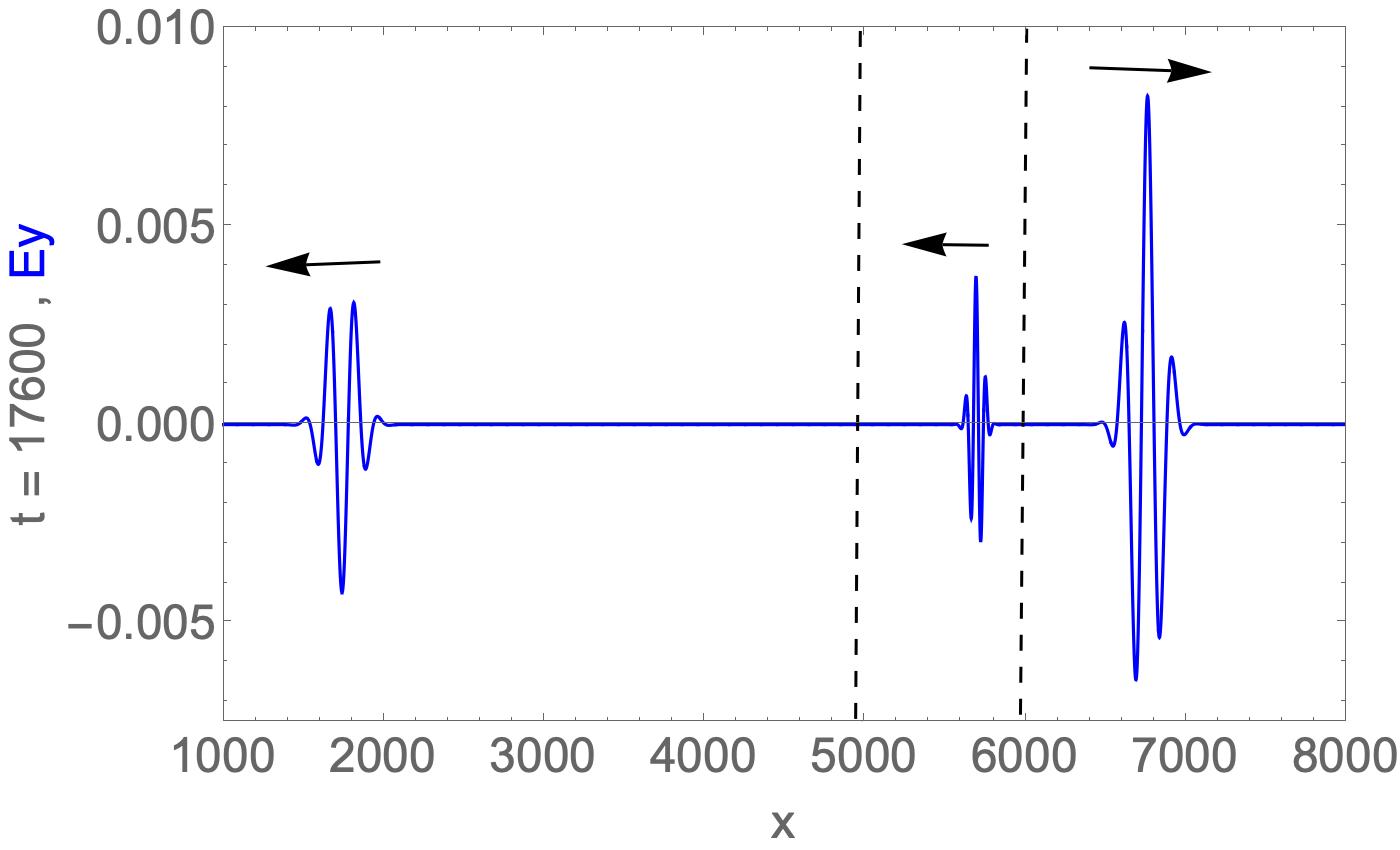} 
\caption{The electric field $E_y$ of a Gaussian wave packet following the second reflection/transmission at the back of the dielectric interface,  $x=6000$.
}
\end{center}
\end{figure}
The asymmetry continues in the transmitted part of the wave packet into the vacuum region, $x > 6000$.

\section{Conclusions and Comments}
\quad  We [1-5] have been developing QLAs for the full Maxwell equations for 1D, 2D and 3D electromagnetic propagation in scalar dielectric media.  While these algorithms utilize quantum information science their actual implementation on error-correcting quantum computers will be decades away.  However, these QLAs are ideally parallelized on classical supercomputers up to the maximum cores available.  Indeed, on the IBM $Mira$ supercomputer, the parallelization of QLAs (for spinor Bose-Einstein condensates [14-16]) still scaled even to over 750 000 cores.  The QLA building blocks are the Khan [7] representation of the full Maxwell equations using the two Riemann-Silberstein-Weber polarization 4-vectors in that medium.  We [1-5] have seen that for both 1D x- and y- propagation one requires a QLA of at least 8 qubits.  For 1D z-propagation,however, the QLA will requre a 16 qubit representation per lattice node.  This is because the QLA now involves the diagonal Pauli spin matrix $\sigma_z$, and since the collision operator must entangle at least 2 qubits at each site the QLA representation requires 16 qubits/node.

An interesting, and somewhat unexpected result of our 1D QLA  pulse simulations [1-2] was that the transmitted to incident electric field amplitude was 
\begin{equation}
\frac{2 n_1}{n_1+n_2} \sqrt{\frac{n_2}{n_1}}
\end{equation}
- a factor of $\sqrt{{n_2}/{n_1}}$ different from the standard boundary value plane wave result [13].
 We [2] then developed a theory on Gaussian pulses that explained this QLA result.  However for an electromagnetic wave packet, the 1D QLA results are much more complicated and one does not recover such a universal result as Eq. (9).

The central point of this note was to develop a 1D QLA for the two curl equations of Maxwell, Eq. (1).  For 1D transverse pulse propagation in a scalar dielectric, the two divergence equations of Maxwell are identically zero - but now, for each polarization and x-propagation, there are only two non-zero field components:  either non-zero $(E_y, B_z)$ or $(-E_z, B_y)$.  Thus only a 4-qubit QLA is needed.  We have here developed such a 4-qubit QLA and performed simulations for both polarizations.  Once again, Eq. (9) is recovered for the ratio of transmitted to incident electric field amplitudes.   For wave packets, there is no such simple universal relationship:  the dominant result is that the wavelenth of the transmitted to incident wave packets scales as $n_1/n_2$.  These simulations were undertaken partly to examine the conference paper by Jestadt et. al. [6].  While they too base their algorithm on the Khan representation for the two curl equations, they directly move to an approximation of the exponential of the sum of two non-commuting operators.  While we have moved to a QLA representation, Jestadt et. al.[6]  applied the Baker-Campbell-Hausdorff approximation.  In their method the final equations involve both $n'(x)$ and $n''(x)$ derivatives of the refractive index.  One unexpected result of their simulations was that the transmitted and incident wave packets had approximately the same wavelengths.  Our 1D QLA 4-qubit simulations yielded wavelength ratios of $n_1/n_2$.  We feel that this could be due to the occurrence of the second derivative in the refractive index, $n''(x)$ in [6].  A little thought had us realize that the first derivative in the refractive index, $n'(x)$, is a term introduced by 
use of the Khan Riemann-Silberstein-Weber vector in the medium, Eq. (2).  Because in the 1D scalar dielectric problem there is no derivative $n'(x)$ introduced when looking at the curl equations of Maxwell, this $n'(x)$ introduced by the material media $\mathbf{F^{\pm}}$ must be counter-balanced by a similar term coming from the introduction of appropriate QLA potential operators.  Similar cancellations most likely occur for the Jestadt representation for the $n'(x)$, but there may not be such a similar cancellation for the $n''(x)$.

Clearly these results indicate that one should actually work with the vacuum Riemman-Silberstein-Weber vectors $\sqrt{\epsilon_0} \mathbf{E} \pm i \mathbf{B}/\sqrt{\mu_0}$ rather than the medium Riemann-Silberstein-Weber vectors  $\sqrt{\epsilon} \mathbf{E} \pm i \mathbf{B}/\sqrt{\mu_0}$.  This will be explored in future research.

\section{Acknowledgments}
This research was partially supported by Department of Energy grants DE-SC0021647, DE-FG02- 91ER-54109, DE-SC0021651, DE-SC0021857, and DE-SC0021653. 

 \section{References}
\quad \, [1]  VAHALA, G, VAHALA, L,  SOE, M $\&$ RAM, A, K.  2020.  Unitary Quantum Lattice Simulations for Maxwell Equations in Vacuum and in Dielectric Media, J. Plasma Phys $\bf{86}$, 905860518 

[2]  RAM, A. K., VAHALA, G., VAHALA, L. $\&$ SOE, M 2021  Reflection and transmission of electromagnetic pulses at a planar dielectric interface - theory and quantum lattice simulations, AIP Advances $\bf{11}$, 105116 

[3]  VAHALA, G, SOE, M, VAHALA, L, $\&$ RAM, A. K.  2021  Two Dimensional Electromagnetic Scattering from Dielectric Objects using Quantum Lattice Algorithm, submitted J. Comp. Physics:  arXiv:2110.05480

[4]  VAHALA, G, VAHALA, L,  SOE, M $\&$ RAM, A, K.  2021  One and Two Dimensional Quantum Lattice Algorithms for Maxwell Equations in Inhomogeneous Scalar Dielectric Media I: Theory,  Radiat. Eff. Defects Solids $\bf{176}$, 49-63

[5]  VAHALA, G, SOE, M,  VAHALA, L, $\&$ RAM, A, K.  2021  One and Two Dimensional Quantum Lattice Algorithms for Maxwell Equations in Inhomogeneous Scalar Dielectric Media I: Theory,  Radiat. Eff. Defects Solids $\bf{176}$, 64-72

[6]  JESTADT, R, APPEL, H, $\&$ RUBIO, A.  2014  Real-time evolution of Maxwell equations in spinor representation, Conference paper.

[7]  KHAN, S. A. 2005  Maxwell Optics:  I.  An exact matrix representation of the Maxwell equations in a medium.  Physica Scripta 71, 440-442;  also arXiv: 0205083v1 (2002)

[8]  LAPORTE, O. $\&$ UHLENBECK, G. E. 1931 Application of spinor analysis to the Maxwell and Dirac  equations. Phys. Rev. 37, 1380-1397.

[9]  OPPENHEIMER, J. R. 1931 Note on light quanta and the electromagnetic field. Phys. Rev. 38, 725-746.

[10]  MOSES, E.  1959  Solutions of Maxwell's equations in terms of a spinor notation:  the direct and inverse problems.  Phys. Rev. 113, 1670-1679

[11]  YEPEZ, J. 2002 An efficient and accurate quantum algorithm for the Dirac equation. arXiv: 0210093. 

[12]  YEPEZ, J.  2005 Relativistic Path Integral as a Lattice-Based Quantum Algorithm. Quant. Info. Proc. 4, 471-509. 

[13]  JACKSON, J, D.  1998. Classical Electrodynamics, 3rd Ed., (Wiley, New York) 

[14]  YEPEZ, J, VAHALA, G, VAHALA, L $\&$ SOE, M.   2009b Superfluid turbulence from quantum Kelvin wave to classical Kolmogorov cascades.  Phys. Rev. Lett. 103, 084501. 

[15]  VAHALA, G, ZHANG, B, YEPEZ, J, VAHALA. L $\&$ SOE, M.  2012 Unitary Qubit Lattice Gas Representation of 2D and 3D Quantum Turbulence.  Chpt. 11 (pp. 239 - 272), in Advanced Fluid Dynamics, ed. H. W. Oh, (InTech Publishers, Croatia)

[16]  VAHALA, G, SOE, M $\&$ VAHALA, L.  2020  Qubit Unitary Lattice Algorithm for Spin-2 Bose Einstein Condensates: II  Vortex Reconnection Simulations and non-Abelian Vortices.  Rad. Eff. Def. Solids 175, 113-119
 
\end{document}